# A Fuzzy Set-based Approach for Matching Hand-Drawing Shapes of Touch-based Gestures for Graphical Passwords

Adel A. Sabour[1,**], Ahmed M. Gadallah[2,*] and Hesham Hefny[3,*]

* Computer Science Department, Institute of Statistical Studies and Research, Cairo University, Egypt.
** Information Systems Department, Institute of Statistical Studies and Research, Cairo University, Egypt.
[1] adelsabour@gmail.com, [2] ahmgad10@yahoo.com and [3] hehefny@ieee.com

**Abstract.** This paper presents a two-dimension fuzzy set based approach for matching touch-based gestures using fuzzy cued click point technique. The proposed approach aims mainly to improve the acceptance of the most closed inaccurate hand drawn gestures generated by the user compared with a predefined referenced gesture value that is stored in the user profile. Commonly, gestures are used in order to facilitate the interactive capabilities between humans and computerized systems. Unfortunately, most of current gesturing techniques don't deal at the same level of inaccuracy of gesturing, resulted from the nature of human fingers and hands movements. This paper aims, in a more flexible manner, to tackle the inaccuracy problem existed with gesture-based interactions between humans and a computerized system.

## 1  Introduction

Human Computer Interaction HCI involves the study, planning and design of the interaction between people and Computers [2] with easiest possible way. Psychological studies explained that human brain recognizes and recalls images and gestures better than text.

### 1.1  Touch-based Gesture

Nowadays, as touch screen devices become common, people rely on using gestures to control devices like smartphones, tablets, laptops, tables and so on [6]. Accordingly, it is very important to propose and develop solutions to handle gestures matching in a more flexible and accurate manner [21]. Generally, hand drawn gestures over a touch screen for example are detected using sensors that return information over only two dimensions X and Y of plane. Consider the Track-pad on a Laptop, It senses a user finger and tracks the movement of such finger. However, sensing is done mainly to gather the movements of the user finger over the two dimensions (X, Y) plane representing the device screen. In other words, touch-screen devices have a gesture recognition systems over a 2D plane to recognize finger movements and , accordingly, hand drawn shapes[3]. In consequent, such recognized hand drawn gestures are translated into appropriate actions and events on the device. Commonly, some interactive operations such as scroll, zoom in and zoom out, draw, etc. can be performed on touch screen based devices using their corresponding gestures [20].

Commonly speaking, hand-drawn shape on a touch screen represents one of the most important types of gestures. It can have one or multiple strokes. Each stroke is a sequence of timed points [14]. Also, gesture can be defined as a non-verbal or non-vocal form of communication in which visible bodily actions communicate particular messages with a computerized system, either in place of, or in conjunction with, speech. Gestures include movement of the hands, face, or other parts of the body [8].

### 1.2 Fuzzy Logic and Fuzzy set theory

Fuzzy logic represents a generalization of crisp logic. It has been proposed mainly for reasoning under uncertainty. It depends mainly on the approximate rather than exact concept of partial truth- truth values between "completely true" and "completely false" [9]. The importance of fuzzy logic derived from the fact that most models of human reasoning and especially common sense reasoning are approximate in nature. A membership function (MF) is a curve that defines how each point in the input space is mapped to a membership value (or degree of membership) between 0 and 1. The input space is sometimes referred to as the universe of discourse, a fancy name for a simple concept [10]. Figure (1) illustrates bivalent sets to characterize a sequence of timed points that create a gesture. Each referenced pixel $rp_i$ has its counterpart pixel $cp_i$ with completely match or un-match at all, regardless the place of counterpart pixel near or far[11].

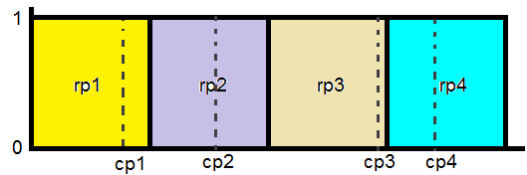

**Fig.1.** Bivalent sets to characterize a sequence of timed points that create a gesture.

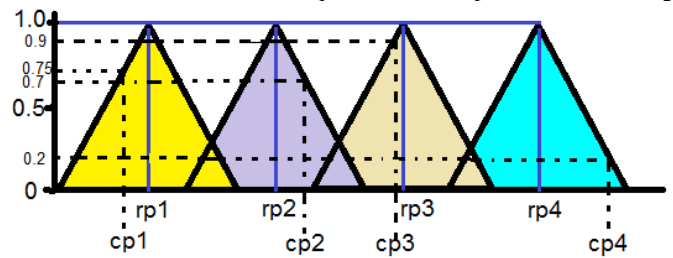

**Fig.2.** Fuzzy sets to characterize a sequence of timed points that create a gesture.

On the other hand, figure 2 illustrates fuzzy sets. Each fuzzy set describes how much a given pixel is closed to the referenced core pixel. Accordingly, an aggregation of the matching degrees of some input pixels and some referenced pixel can be used to judge how much the input pixels are closed to the referenced ones. Each referenced pixel $rp_i$ has its counterpart pixel $cp_i$ with a matching degree. For example cp1 has 0.75 as degree of matching rp1, and cp4 has 0.2 as degree of matching rp4. That illustrates the degree of matching counterpart pixel to its referenced pixel. In fact both gesture values are similar at least for some extent although they are not completely identical [12].

The rest of this paper is organized as follows: section 2 presents the related work. The problem of traditional crisp approaches is addressed in Section 3. Section 4 shows The proposed Fuzzy Cued Click-Points (FCCP) Approach. The proposed fuzzy based approach for gesture values matching is addressed in Section 5. Section 6shows an illustrative case study. The conclusion is given in section 7.

## 2  Related Work

There are many different techniques that have been used for matching clicks values. In **pass-points** technique [15, 16], passwords consist of a sequence of five click-points on a given image. Users may select any pixels in the image as click-points for their password. To log in, they repeat the sequence of clicks in the correct order. Pass-Points uses a 20x20 pixel tolerance square of the original click-point. On the other hand, **CCP- Cued Click-Points** technique uses one click-point on five different images in sequence rather than 5 clicks in single image [17]. The next image displayed is based on the location of the previously entered click-point, creating a path through an image set, within a system-defined tolerance distance of the original click-points.

In **persuasive cued click-points (*PCCP*)** [18], the theoretical password space is ($w$ x $h/t^2$)$^C$, where the size of the image in pixels ($w$ x $h$) is divided by the size of a tolerance square ($t^2$, in PCCP experiments, $19^2$), to get the total number of tolerance squares per image raised to the power of the number of click-points in a password (c, usually set to 5 in PCCP experiments). Also, an **enhanced cued click-points (ECCP)** [19], is initially functions like cued click points. But during a password creation, a discretization method is used to determine a click-point's tolerance square and the corresponding grid.

Commonly, there are many approaches that have been used for matching gesture values respecting traditional two-valued logic as presented in the next section.

**Bounding box method** involves drawing a box around the gesture value and dividing it up into a grid [5]. The gesture is then defined by the areas that it passes through. The grid would be set up as shown in fig.5 where the gesture on the right would then be parsed as 13,9,5,1,2,3,4,8,12 and 16 if you had drawn it from left to right.

**Fig.3.**Bounding box method [5]

The advantage of **corner detection method** is that it would be very accurate, the provided algorithm detects the corners properly. This method also takes into account the proportions of each part of the gesture. In this method we determine the top left and bottom right corners of the bounding box. We use this method because it uniquely represents every user selected regions in the used image, that is, no two different regions on the image will have the same Corner value [5].

**Region Pixel Count Method** includes figuring out which points are the corners, after that looking at the relationship between those corners. It specifies how many virtual grid pixels the user had used to choose the region of gesture. Two different user may choose the same region in the same image but their selection cannot be the same. Hence they differ on the number of the virtual pixels and also it distinguishes the same region selection with different shape because the two shape will have different pixel sums.

Generally, the **start and end points of gestures** are based on the direction that calculated by the competitive differential observation probability value, which is determined by the difference of observation probability value of maximal gesture models and non-gesture model. The key gesture starts (ends) when the value of competitive differential observation probability changes from negative to positive (positive to negative). To reduce the states of the non-gesture model, model reduction which merges similar probability distributions states based on relative entropy is used [3].

## 3      The problem of traditional crisp approaches

Gestures values by nature are imprecise approximated values, i.e. the user uses a hand to generate a gesture value that approximately matches his/her predefined one.

Consequently and despite facilitating the user interaction for the authentication operation using a touch screen, a complex problem of matching user approximated gesture values has been appeared [4]. For, example when a user draw a circle at a specific location over a displayed image and repeat the operation at different time, he/she tries to make the second gesture value as closed as possible to the first one. In fact both gesture values are similar at least for some extent although they are not completely identical.

As a gesture value has a set of pixels, two gesture values matching process involves evaluating the matching between each two counterpart pixels. Traditionally, for matching two gesture values, most of the current approaches give an accepted range or tolerance distance as a radius of a circle centered on the referenced pixel in the first gesture value. Then, each counterpart pixel in the second gesture value must belong to the accepted range within such circle of its referenced pixel else there is no matching between the two gesture values at all.

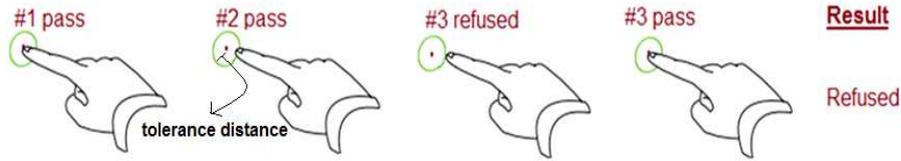

**Fig.4.** A mismatching 4-pixels two gesture values because of mismatching of the counterpart pixel no 3 with its referenced pixel at the circle center.

As an example, Fig. 3 shows a traditional approach [9, 15, 16, 17, 18, and 19] for matching 4 counterpart pixels as a simplification of a gesture idea. Each referenced pixel rpi is the center of a circle representing the matching domain of its counterpart pixel cpi. Accordingly, a two pixels are matching if and only if :cpi∈dom(rpi). Consequently, although three pixels of the entered four pixels are matching with their referenced pixels, the two gesture values are mismatching because of the third pixel that is not matching with its referenced pixel using Eq. (1) as shown in Fig. (3).

$$Mf(rp_i, cp_i) = \begin{cases} 1 & cp_i \ni dom(rp_i) \\ 0 & cp_i \not\ni dom(rp_i) \end{cases} \quad \text{eq.(1)}$$

## 4  The proposed Fuzzy Cued Click-Points (FCCP) Approach

This work proposes an enhanced two-dimensional fuzzy set approach in the process of matching two pixels and deterring how much they are closed to each other. Such approach determines in a flexible manner how much a pixel values (X, Y) matches another referenced one with a matching degree in the range [0, 1]. Commonly, two pixels values are considered to be identical if their matching degree is 1. On the other hand, 0 matching degree indicates that two pixels values are considered to be mismatching at all. In consequent, a matching degree greater than 0 and less than 1 between two gesture values indicates a partial matching between such two pixels values. In other words, a matching degree between two pixels values stats clearly how much both of them are identical or closed to each other. Consequently, a threshold value must be defined that determines which matching values are considered accepted.

For example, in a fuzzy cued click-points approach the user determines a threshold value indicating which entered gesture is accepted and which is rejected. As shown in Fig. (4) as an , the 4-pixel two gesture values match by 55% as a result of partial matching of some of their counterpart pixels. Essentially, a matching degree between two pixels represents how much they are closed to each other. Actually, the matching by 55% of 4-pixel two gesture values is computed as the average of the matching degrees between each referenced or base pixel and its counterpart or corresponding pixel as ((1+0.2+0+1)/4). Intuitively, Fuzzy membership functions respecting two-dimension universe of discourse are very suitable and efficient to compute such matching degrees as presented in the fifth section. Thus, the proposed two-dimensional fuzzy set approach to gesture values matching makes it more flexible and human-like.

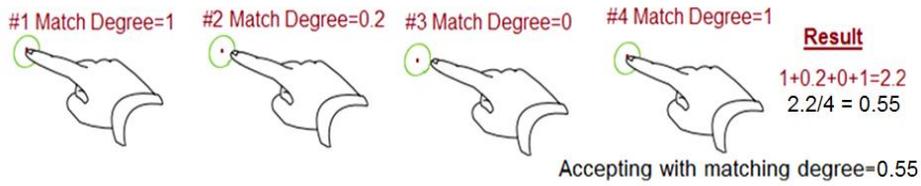

**Fig.5.** Two gesture values that match partially by 55%

As explained, in traditional crisp approaches of matching two gesture values, a mismatching of two counterpart pixels leads to a mismatching of the two gesture values at all. In contrary, the proposed fuzzy based approach determines how much two gesture values are matching in a human-like fashion.

## 5   The Proposed 2D Fuzzy Set Based Approach

The proposed two-dimension fuzzy set based approach depends on fuzzy cued clickpoints (FCCP) technique and defining the gesture as sequence of timed points or pixels [13]. Generally, each referenced pixel $rp_i$ is represented by an ordered pair ($x_{rpi}$,$y_{rpi}$) over a two dimensional space with X and Y axis [1]. Also, each counterpart pixel $cp_i$ is represented by another ordered pair ($x_{cpi}$,$y_{cpi}$) over the two dimensional space representing the underlined surface of the touch screen. Accordingly, a membership function for matching two counterpart pixels can be as shown in eq. (2) and fig. 6. On the other hand, eq. (3) computes the matching degree between two gesture values computed as the average of matching each counterpart and referenced pixels as shown in eq. (3).

$$\mu_{match}(rp_i, cp_i) = \mu_{around}(x_{rp_i}, x_{cp_i}) \text{ AND } \mu_{around}(y_{rp_i}, y_{cp_i}) \qquad \text{eq.(2)}$$

$$m(g_1, g_2) = \sum_{i=1}^{n} \mu_{match}(rp_i, cp_i)/n \qquad \text{eq.(3)}$$

where g1 and g2 represent two gesture values, $rp_i$ and $cp_i$ represent each referenced and its counterpart pixel in the two gesture values.

Fig.6 demonstrates the idea of the proposed two-dimension fuzzy set approach in gestures matching. Such fuzzy set can be used to define the fuzzy number 'around' over a two dimensional universe of discourse. Commonly, the fuzzy number 'around' is mainly used to measure the matching degree of how much a point $p_i$, represented as ordered pair $(x_{pi}, y_{pi})$, closed to or around another point $p_j$ represented as ordered pair $(x_{pj}, y_{pj})$.

When a user generates a gesture value using his/her finger or mouse pointer to an image displayed ona screen, the resulting gesture value is considered a series of points. Each point has two values of two dimensions X and Y. The dimensions of any point are measured by how it is far from the start point of the horizontal dimension X and how it is far from the start point of the vertical dimension Y. Accordingly, a point(X, Y) represents the point values over the two dimensions.

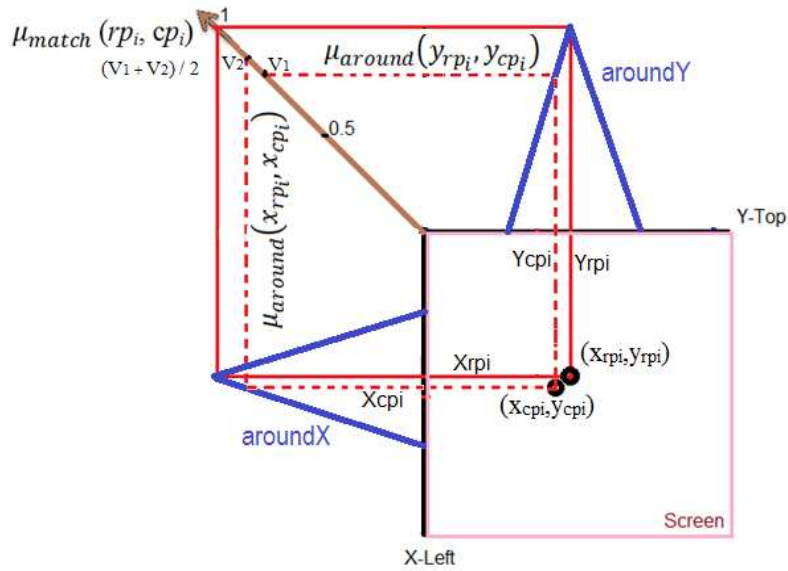

**Fig.6.** The proposed two-dimension fuzzy set approach

In gesture based password, when a user finger or a mouse pointer generates a gesture value as a password, then a series of pixels are stored in the user profile database. Each password pixel, called a referenced pixel, is stored as an ordered pair $(x_{rpi}, y_{rpi})$. Consequently, when a user aims to login, he should enter a more closed gesture value to his/her stored password. Consequently, each pixel in the entered password, called a counterpart pixel $(x_{cpi}, y_{cpi})$, is checked to measure how much it matches its referenced pixel $(x_{rpi}, y_{rpi})$.

In the proposed approach, such matching operation is basically based on the idea of two-dimension fuzzy set and by applying eq. (2). Reasonably, the resulted membership value represents how much the counterpart pixel matches the referenced one. In consequence, the average of matching degrees, between each counterpart pixel and its referenced pixel, represents the overall matching degree of the entered gesture with the referenced stored gesture. As the user set an accepted threshold, a user valid gesture if the overall matching degree is greater than or equal the accepted threshold else the matching fails.

## 6  Illustrative Case Study

As an application of the proposed approach, a user attempts to generate a gesture password in a subscription process for a computerized system. Firstly, the user should upload an image to draw his/her specific gesture password shape[7]. Such, entered gesture value is stored in the user profile in the database as a set of pixels of ordered pairs. Obviously, each pixel is stored in the database as two values over horizontal and vertical dimensions of the used image in password generation. Also, as complementary step, the user should determine an acceptance threshold value. Accordingly, a login process succeeds if the matching degree between a login password and the user stored password is greater than or equal the set acceptance threshold value.

Consequently, when the user needs to logon to the system, he loads his/her password image and tries to draw a gesture shape as closed as possible to his/her saved gesture password.

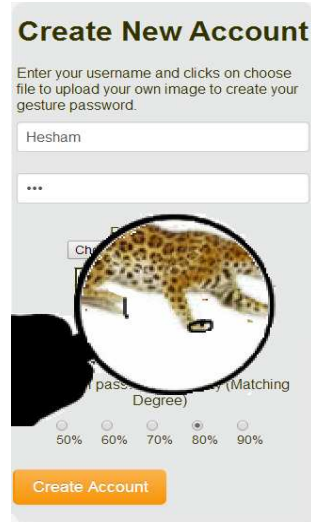

**Fig.7.** A user set a gesture value password with 0.8 acceptance threshold.

Fig. 7 shows a screenshot of the gesture based password application. Interaction with the application to create a user account with a gesture value password can be described as follows:

- The user creates a new account with a user name then uploads an image and draws his/her gesture password value that will be stored in the user profile as a set of pixels.
- The user sets an acceptance threshold value reflecting the degree of gesture accuracy.
- The user click on create account store the entered user account values in the user profile.

On the other hand, an interaction with the application to logon to the system can be described as follows:

- The user enters the Username. Then, the image used with the user password is displayed.
-  The user draws a gesture shape as closed as possible to his/her stored password and clicks the logon button.
- A process of matching the login entered gesture value with the user stored gesture value password will take place. The result of the matching process is a matching degree reflecting how much the login password is closed to the user stored password.
- Finally, the user reaches a successful login if the resulted matching degree is greater than the user predefined accepted threshold, as shown in fig. 8. Otherwise, the login fails, as shown in fig. 9.

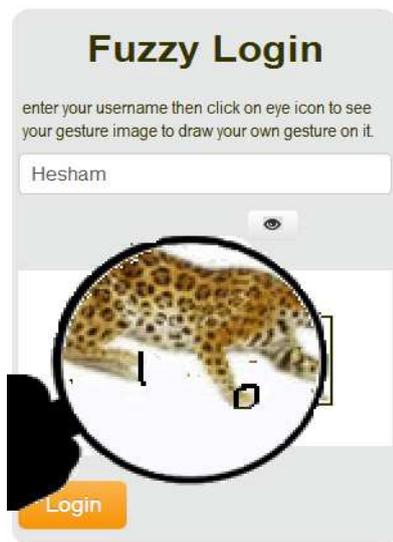
**Fig. 8.** Successful logging with a gesture value password with 90% matching degree

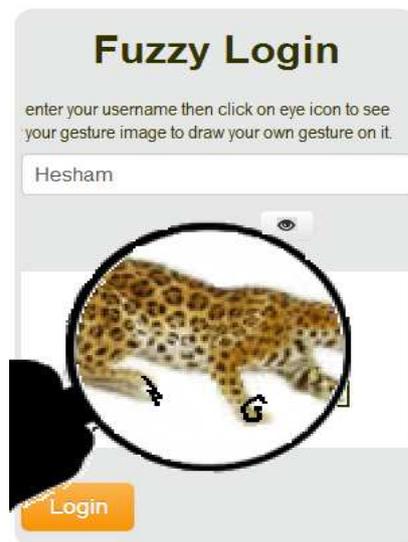
**Fig. 9.** Failed logging with a gesture value password with 74% matching degree

Almost, an entered gesture value, as a login password, is shifted vertically or horizontally regarding the referenced user stored password. Such case may lead to rejecting the checked gesture password although it may be greatly identical to the user stored password. To solve this problem, the first pixel of the saved gesture password is used to find the more closed pixel in the leading pixels of the entered login gesture password or vice-versa. As a result, we reach the more closes counterpart pixel to the first referenced pixel in the stored gesture password.

# 7    Conclusion

Many approaches have been proposed for matching gesture values based on the two valued traditional logic. In such approaches two gesture values are completely matching or completely mismatching. In reality, two much closed hand drawn gesture values are matching partially with a matching degree between completely mismatching and completely matching. This paper presents a novel fuzzy based approach for gestures matching in a novel manner. The proposed approach depends mainly on the idea of two dimension fuzzy set representing the width and the height of the image used as a background for the gesture value. According to the proposed approach, two hand drawn gesture values are matching if the average matching degree of each referenced pixel and its counterpart pixel is greater than a predefined accepted threshold. The proposed approach is applied to a user login application in which a user draws a gesture value over a selected image as a login password. Consequently, the user reaches a successful login if the resulted matching degree is greater than the user predefined accepted threshold. Otherwise, the login fails. Accordingly, the proposed fuzzy based approach matches gesture values in a more flexible and human-like manner.

## References

1. Joao P. et al., "Two-Input Fuzzy TPE Systems", Portuguese Foundation for Science and Technology, 2004.
2. Mohanbhai C., "Tool Interaction and Information Processing: HCI Evolution", International Journal of Application or Innovation in Engineering & Management, Vol. 3, Issue 1, January 2014, ISSN: 2319–4847.
3. Pratibha T. et al., "Human Computer Interaction: Analysis and Journey through Eras", International Journal of Computer Science and Mobile Computing, Vol. 3, Issue. 4, April 2014, pp. 653–659.
4. Peralta R. T. et al.,"Interactive Gesture-based Authentication for Tabletop Devices", Report, November 2011.
5. Niranjana G. and Dawan K. , "Graphical Authentication Using Region Based Graphical Password", International Journal of Computer Science and Informatics, 2231 –5292, Vol. 2, Issue 3, pp.6-11, 2012.
6. Feng T. et al., "Continuous Remote Mobile Identity Management Using Biometric Integrated Touch-Display", 45th Annual IEEE/ACM International Symposium on Microarchitecture Workshops, 2012.
7. Fong S. et al., "A biometric authentication model using hand gesture images", BioMedical engineering, 2013.
8. Adam K., Gesture: Visible Action as Utterance. Cambridge: Cambridge University Press, 2004, ISBN: 0-521-83525-9.
9. Sabour A. et al., Flexible Querying of Relational Databases: Fuzzy Set Based Approach, CCIS 488, pp. 446–455, 2014.
10. Dietz R. and Moruzzi S. (eds.), Cuts and clouds. Vagueness, Its Nature, and Its Logic, Oxford University Press, 2009.
11. Ann Markusen, "Fuzzy Concepts, Scanty Evidence, Policy Distance: The Case for Rigour and Policy Relevance in Critical Regional Studies." In: Regional Studies, Vol. 37, Issue 6-7, 2003, pp. 701-717.
12. Malhotra R et al., "Self-adaptive Gesture Classifier Using Fuzzy Classifiers with Entropy Based Rule Prunning", Intelligent Informatics, AISC 182, pp.217-223, 2013.

13. Vivek V. J. and Swaminathan P. L. ,"Robust Hand Gesture Recognition Algorithm for Simple Mouse Control", International Journal of Computer and Communication Engineering, Vol. 2, No. 2, March 2013.
14. "http://developer.android.com/reference/android/gesture/Gesture.html", Octobar, 1, 2014.
15. S.Wiedenbeck, J.Waters, J. Birget, A. Brodskiy, and N. Memon, "PassPoints: Design and longitudinal evaluation of a graphical password system," International Journal of Human - Computer Studies, Vol. 63, No. 1-2, pp.102–127, 2005.
16. Dirik, A.E., N. Menon, and J.C Birget. Modeling user choice in the PassPoints graphical password scheme. ACM SOUPS, July18-20, 2007.
17. S. Chiasson, P. van Oorschot, and R. Biddle, "Graphical password authentication using Cued Click Points," in European Symposium On Research In Computer Security (ESORICS), LNCS 4734, September 2007, pp. 359–374.
18. Sonia Chiasson, Elizabeth Stobert, Alain Forget, Robert Biddle, P. C. van Oorschot, "Persuasive Cued Click Points: Design, implementation, and evaluation of a knowledge-based authentication mechanism", IEEE Trans, Vol. 9, Issue 2, 2011.
19. Lavanya Reddy L and K.Alluraiah, "ECCP: Enhanced Cued Click Point Method for Graphical Password Authentication", International Journal of Advanced Research in Computer Science and Software Engineering, Volume 3, Issue 8, August 2013.pp. 321-325.
20. Jolly A. et al. "Contactless Gesture Recognition System Using Proximity Sensors", International Journal of Advanced Research in Computer Science and Software Engineeringijarcsse, Vol. 4, Issue 2, pp. 824-828, 2014.
21. MENG Y. "Gestures Interaction Analysis and Design Strategy", Master Thesis Report University of Birmingham, 2014.